\documentclass[preprint2]{aastex} 

\shorttitle{Millisecond variability in SAX J1750.8-2900}
\shortauthors{Kaaret et al.}

\begin{document}

\title{Discovery of Millisecond Variability in the Neutron-Star X-Ray
Transient SAX J1750.8-2900}

\author{P.\ Kaaret} \affil{Harvard-Smithsonian Center for Astrophysics,
60 Garden St., Cambridge, MA 02138, USA}
\email{pkaaret@cfa.harvard.edu}

\author{J.J.M.\ in 't Zand} \affil{Astronomical Institute, Utrecht
University, P.O. Box 80 000, 3508 TA Utrecht, the Netherlands \\ SRON
National Institute for Space Research, Sorbonnelaan 2, 3584 CA Utrecht,
the Netherlands}

\author{J.\ Heise} \affil{SRON National Institute for Space Research,
Sorbonnelaan 2, 3584 CA Utrecht, the Netherlands}

\author{J.A.\ Tomsick} \affil{Center for Astrophysics and Space
Sciences, Code 0424, 9500 Gilman Drive, \\ University of California at
San Diego, La Jolla, CA 92093}

\begin{abstract}

We report the discovery of millisecond oscillations in the X-ray
emission from the X-ray transient SAX J1750.8-2900.  Millisecond
quasiperiodic oscillations (kHz QPOs) were present in the persistent
emission with frequencies ranging from 543~Hz to 1017~Hz.  Oscillations
at a frequency of 600.75~Hz were present in the brightest X-ray burst
observed.   We derive an upper limit on the source distance of $6.3 \pm
0.7 \rm \, kpc$ from this X-ray burst.  

\end{abstract}


\section{Introduction}

The discovery of millisecond oscillations in the persistent emission
and also in thermonuclear X-ray bursts from neutron star low-mass X-ray
binaries (LMXBs) has opened a new window on the dynamics of accreting
neutron stars \citep{strohmayer96}.  The time scales of the kHz
quasiperiodic oscillations found in the persistent emission (hereafter
kHz QPOs) match those expected from accretion dynamics very close to
the neutron star in a region of strong gravity, and the possibility
that the kHz QPOs could be used as probes of strong gravity is very
intriguing.  The burst oscillations are in a range expected for the
neutron star spin periods \citep{alpar85}, and have been interpreted as
evidence for millisecond spin periods in accreting neutron stars
\citep{strohmayer97}.  However, this interpretation has recently been
questioned because of the large frequency shifts seen in some bursts.
Observations of bursts from new sources would be useful in determining
the correct interpretation of the burst oscillations.

Here, we describe observations made with the Rossi X-Ray Timing
Explorer (RXTE; Bradt, Rothschild, \& Swank 1993) following the
detection of X-ray bursts from the transient source SAX J1750.8-2900. 
We report the discovery of  kHz QPOs in the persistent emission and of
oscillations in one X-ray burst.  We describe the source and our
observations in \S 2, our results on X-ray bursts in \S 3, our results
on the persistent emission in \S 4, and discuss our results in \S 5.

\section{Observations of SAX J1750.8-2900}

SAX J1750.8-2900 was discovered in March 1997 with the Wide Field
Cameras (WFCs) on BeppoSAX as a faint and short duration transient
1\fdg 2 from the  galactic center showing type-I X-ray burst activity
\citep{bazzano97}.  Analysis of data from the WFCs and the All-Sky
Monitor (ASM) on RXTE shows a peak flux of 0.12~Crab (where the unit
`Crab' is defined as the flux of the Crab nebula in the 2-10~keV band,
$\sim 2 \times 10^{-8} \rm \, erg \, cm^{-2} \, s^{-1}$) and activity
lasting approximately 3 weeks above 0.01~Crab \citep{natalucci99}. The
WFCs detected nine type-I X-ray bursts, with peak fluxes between 0.4
and 1.0~Crab and $e$-folding decay times between 4 and 11~s in the
2-8~keV band.  No counterparts in other wavelength regimes were
identified. The type-I bursts indicate that the source is an accreting
neutron star.

\begin{figure}[tb] \epsscale{1.0} \plotone{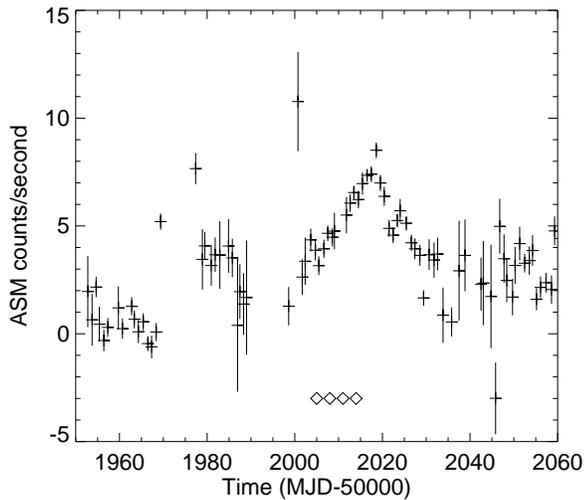} \caption{RXTE/ASM
light curve of the 2001 outburst of SAX J1750.8-2900.  The points are
one day averages of the ASM count rates.  The diamonds indicate the
times of the PCA observations.} \label{asmlc} \end{figure}

SAX J1750.8-2900 turned on again four years later, on 1 March 2001 (MJD
51969), as shown by observations with the RXTE All-Sky Monitor (ASM)
(announced by the MIT ASM Team), the Proportional Counter Array (C.\
Markwardt, private communication), and the WFCs.  The ASM light curve
of the 2001 outburst is shown in Fig.~\ref{asmlc}.  The ASM gives
relatively poor coverage near the beginning of the outburst, so
detailed information is not available on the initial rise.  The high
rate point at MJD 52000 is from a single ASM dwell and may be due to an
X-ray burst.  After the initial outburst, the source flux decayed, and
then rose again with an almost linear dependence of ASM count rate with
time over a period of roughly 20 days to a peak flux of 0.12~Crab,
nearly identical to the 1997 maximum flux.  The source subsequently
decayed over 10 days to a state of reduced flux, well above the level
associated with quiescence in neutron star LMXBs \citep{asai98}.

Bursting activity was detected in the 2001 outburst with the WFCs,
although not as extensively as during the 1997 outburst probably
because the coverage was not as large as in 1997. Nevertheless, the
WFCs detected 4 bursts in a single observation on 4 April 2001 (MJD
52003).  These four bursts have peak fluxes and light curves similar to
the bursts observed in 1997.  Three other bursts were detected with the
WFCs during the 2001 outburst, one 13 days earlier which is atypical
(Natalucci et al.\, in preparation), one 162 days later, and one 179
days later.

The four X-ray bursts found with the WFC triggered an RXTE
Target-of-Opportunity program which led to 4 observations, on 2001
April 6, 9, 12, and 15.  All of our observations occurred during the
second rise in flux, as indicated in Fig.~\ref{asmlc}.  Data were
obtained with the Proportional Counter Array (PCA) in a spectral mode
(Standard 2) with 256 energy channels and 16~s time resolution, a low
resolution timing mode (Standard 1) with no energy information and
0.125~s time resolution, and a high resolution timing mode (event mode)
with 122~$\mu$s time resolution and 64 energy channels.  In addition,
burst catcher modes were used to acquire high time resolution data when
the event rates exceeded the telemetry capacity of the event mode
during X-ray bursts.

\begin{table}[tb]
\begin{center}
\begin{tabular}{ccc}
\tableline
Number & Time & Peak flux \\
\tableline
1 & Apr 06 at 13:26:13 & $6.1 \pm 0.6$ \\
2 & Apr 12 at 14:20:30 & $6.4 \pm 0.6$ \\
3 & Apr 15 at 17:02:25 & $4.8 \pm 0.5$ \\
4 & Apr 15 at 18:37:08 & $1.0 \pm 0.2$ \\
\tableline 
\end{tabular}

\caption{Properties of X-Ray Bursts.  The table gives the time (UTC) at
the start of each burst, and the bolometric peak flux in units of
$10^{-8} \rm \, erg \, cm^{-2} \, s^{-1}$. \label{bursttable}} 
\end{center} \end{table}

\begin{figure}[tb] \epsscale{0.5} \plotone{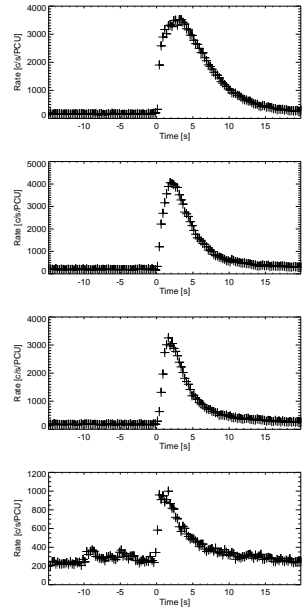} \caption{Light
curves of the four bursts (burst 1 is at the top).  Shown are the PCA
count rates in 0.25~s intervals. The zero times are the burst start
times given in Table~\ref{bursttable}} \label{blcs} \end{figure}

\begin{figure}[tb] \epsscale{0.5} \plotone{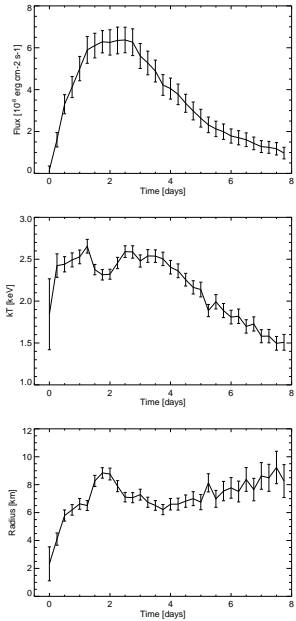} \caption{Spectral
evolution of burst 2.  Shown are the unabsorbed bolometric flux (top
panel), the blackbody temperature (middle panel), and the equivalent
radius assuming a distance of 6.3~kpc (bottom panel).} \label{bur2spec}
\end{figure}

\begin{figure}[tb] \epsscale{1.0} \plotone{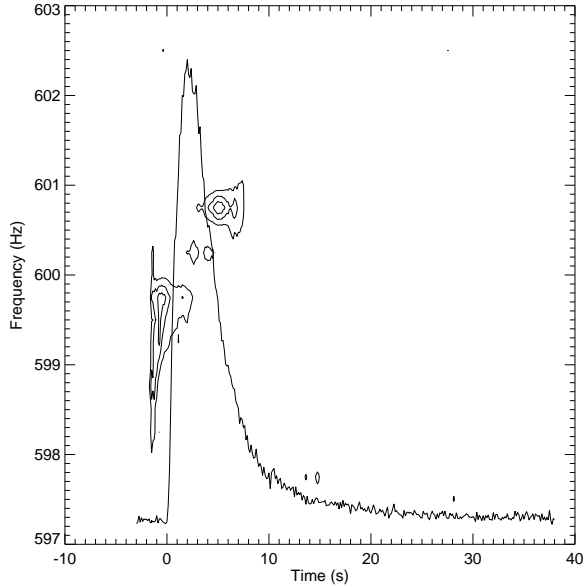} \caption{Dynamical
power spectrum of burst 2 with the burst light curve (count rate)
superimposed.  The contours are at Leahy powers of 12, 24, and 36.  The
contours are generated from power spectra for overlapping 4~s intervals
of data with the power plotted at the mid-point of the interval.  The
power apparently preceding the burst is actually due to the part of the
4~s interval overlapping the burst.} \label{bur2pow} \end{figure}

\section{X-Ray Bursts}

We used the Standard 1 data to search for X-ray bursts and found four
bursts, see Table~\ref{bursttable} and Fig.~\ref{blcs}.   We examined
the spectral evolution of the bursts using burst catcher or event mode
data with 64 channels of energy resolution.  We extracted spectra for
0.25~s or 1~s intervals using all Proportional Counter Units (PCUs) on
during each burst and all layers.  The fluxes were corrected for
deadtime effects.  The maximum deadtime correction in the brightest
burst was 7\%.  In order to eliminate the contribution of the
persistent emission, we subtracted off a spectrum from 10~s of data
preceding each burst, and fit the resulting spectra in the 3-20~keV
band with an absorbed blackbody model with the column density fixed to
the average value found from the persistent emission spectra described
below.

The first three bursts are similar with comparable peak fluxes.  The
decay ($e$-folding) times for the full band (2--60~keV) PCA count rate
for bursts 2 and 3 were close to 3.0~s.  Burst 1 had a broader peak
than bursts 2 and 3, had a longer decay time of 5~s, and deviated
significantly from an exponential decay.  Burst 4 was preceded by a
small precursor event, beginning 9~s earlier with a peak count rate
(with persistent emission subtracted) of $\sim 20\%$ of the main
burst.  Substructure including at least two small peaks is present in
the precursor event.  The peak count rate of burst 4 was a factor of 4
lower than the other bursts and the corresponding temperature was
lower.  Its decay time was 3.4~s, close to that of bursts 2 and 3.

The results of the spectral fits for the brightest burst, number 2, are
shown in Fig.~\ref{bur2spec}.  The burst shows an increase in radius
and a simultaneous decrease in temperature near the peak, but the
magnitude of the radius increase is not sufficient to classify the
burst as an Eddington-limited photospheric radius expansion burst. 
\citet{lewin95} found that the average peak luminosity for photospheric
radius expansion bursts from 5 bursters located in globular clusters,
hence with known distances, was $(3.0 \pm 0.6) \times 10^{38} \rm \,
erg \, s^{-1}$.  Using this luminosity as an upper limit on the peak
luminosity in burst 2, we place an upper limit on the distance to SAX
J1750.8-2900 of $6.3 \pm 0.7 \rm \, kpc$.  This upper limit is
consistent with the upper limit derived by \citet{natalucci99} of
7~kpc.

Using high time resolution data (merged event lists from the event and
burst catcher modes) with no energy selection, we computed power
spectra for overlapping 4~s intervals  of data, with 0.125~s between
the starts of successive intervals, and searched for excess power in
the range 200-1200~Hz.  We found oscillations in the second burst with
a maximum Leahy normalized power of 49.3 at a frequency of 600.75~Hz,
occurring in the burst decay 5~s after the burst rise.  Allowing $2
\times 10^{4}$ independent trials over the 20~s duration of the burst,
the chance probability of occurrence is $4 \times 10^{-7}$, equivalent
to a $5.0 \, \sigma$ detection.

The dynamical power spectrum is shown in Fig.~\ref{bur2pow}. In
addition to the oscillations in the burst decay, there are oscillations
present near 599.5~Hz in the initial part of the burst rise.  The
oscillations begin at the burst onset; no oscillations are present in
the persistent emission before the burst onset.  The frequency appears
to change rapidly between burst rise and decay.

\section{Persistent emission}

To study the persistent emission, we used the Standard-2 data for
spectral information and a 122~$\mu$s time resolution event mode for
timing information.  We removed data around the X-ray bursts.  We used
only PCU 2 for the color and spectral studies as it was on during all
of the observations and has a reasonably well understood response (in
particular, it does not suffer from the propane layer leak experienced
by PCU 0 which was the only other PCU on during all of the
observations).  The background estimate was made using the bright
source background files and the response was calculated using pcarmf
v7.10 as supplied in ftools v5.1.  All PCUs on during each observation
were used for the timing analysis.

\begin{figure}[tb] \epsscale{0.5} \plotone{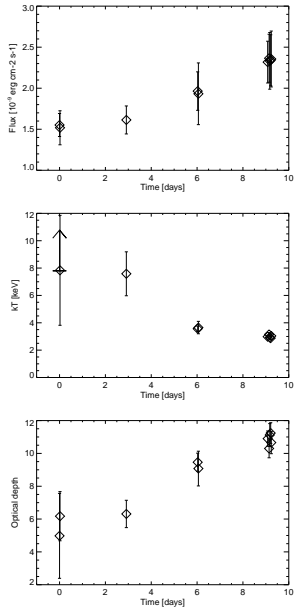} \caption{Spectral
evolution of the persistent emission.  Shown are the absorbed flux in
the 3-20~keV band (top panel), the Comptonization temperature (middle
panel), and the Comptonization optical depth (bottom panel).}
\label{pers_spec} \end{figure}

\subsection{Energy spectra}

We extracted an energy spectrum for each uninterrupted RXTE observing
window to investigate evolution of the energy spectrum.  The energy
range was limited to 3-20~keV and we included a 1\% systematic error. 
We attempted to fit the spectra with a simple absorbed powerlaw model. 
This was unacceptable in all cases.  Addition of a gaussian emission
line at an energy near the Fe-K transition produced an acceptable fit
for the first spectrum, but not for the others.   The sum of a
blackbody plus a gaussian with absorption from material with solar
abundances and the sum of a multicolor disk blackbody plus a gaussian
with absorption produced unacceptable fits.  The sum of a
Comptonization model \citep{st80} and a gaussian emission line with
absorption produced acceptable fits, $\chi^2_{\nu} < 1$, to all of the
spectra.  We used this model to fit all of the persistent emission
spectra.  The limiting form of the Comptonization spectrum at high
temperatures over the fitted energy band is indistinguishable from a
powerlaw given our limited statistics.  Therefore, only an upper limit
on the Comptonization temperature could be obtained for the first
observation which had a spectrum consistent with the sum of a powerlaw
and emission line with absorption as noted above.

The Comptonization temperature and optical depth show marked evolution
over the observations, see Fig.~\ref{pers_spec}.  The temperature
begins high, the lower limit on the temperature in the first
observation is 8~keV, and decreases to 3~keV.  Simultaneously, the
optical depth begins at 6 and increases to 11 for the later
observations.  The column density of the absorbing material shows some
evidence for a decrease with time; however, we caution that these
spectra are not adequate to robustly constraint the column density. 
The average equivalent Hydrogen column density is $4.4 \times 10^{22}
\rm \, cm^{-2}$.

The iron line parameters remain roughly constant across all the
observations but are poorly constrained.  All of the spectra were
consistent with a centroid energy of 6.7~keV and this centroid energy
was subsequently fixed for all of the fits.  The width and
normalization were allowed to vary.  The best fit line widths were in
the range 0.5--1.0~keV.  The best fit equivalent widths ranged from 150
to 400~eV.  The line width is consistent with previous results from
neutron-star LMXBs \citep{white86,asai00}. The equivalent width is
higher than most previously reported values for neutron-star LMXBs; the
maximum equivalent width reported from ASCA is 170~eV \citep{asai00}.

\begin{figure}[tb] \epsscale{1.0} \plotone{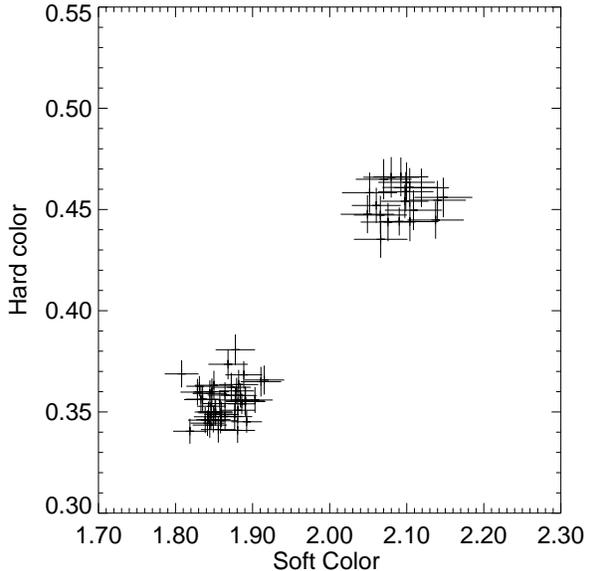} \caption{Color-color
diagram of the persistent emission.  Each point represents 256~s of
data.  The hard color is defined as the count rate in the 9.7--16~keV
band divided  that in the 6.4--9.7~keV band; the soft color is the 
4-6.4~keV band rate divided by the 2.6--4.0~keV band rate.} \label{ccd}
\end{figure}

\begin{figure}[tb] \epsscale{1.0} \plotone{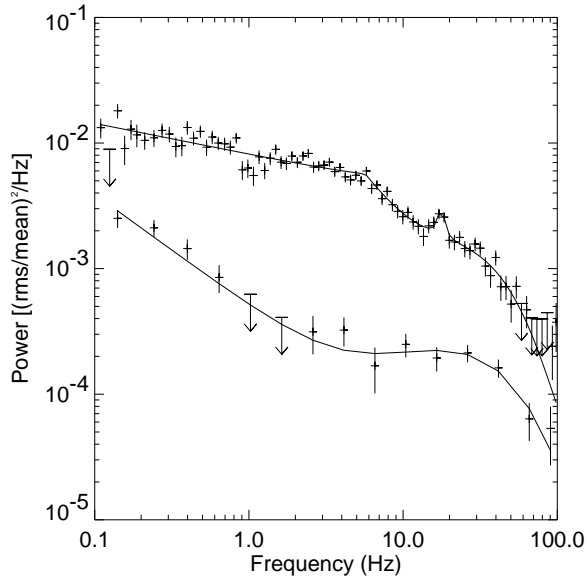} \caption{Low
frequency power spectra.  The upper curve is the power spectrum of the
observations on April 6 and 9.  The  lower curve is for data from April
12 and 15.  The power is RMS normalized.  The solid lines are fitted
curves described in the text.} \label{lowfreq} \end{figure}
\subsection{Source state and classification}

We produced a color-color diagram, see Fig.~\ref{ccd}, using 256~s
intervals extracted from background subtracted Standard-2 data to study
the source state.  The April 6 and 9 observations form one cluster of
points in the color-color diagram with high values of the hard color,
while the April 12 and 15 data form a distinct cluster with lower
values of the hard color.  

To further investigation the source state, we produced low frequency
(below 100~Hz) power spectra.  We produced two spectra, one for each of
the two clusters of points in the color-color diagram, see
Fig.~\ref{lowfreq}.   

The power spectrum for the April 6 and 9 observations shows strong
band-limited noise, with an RMS fraction of 12.8\% in the 0.1--100~Hz
range.  This clearly identifies the source state as the `island' state
\citep{vdk95}.  The power spectrum has the form of a broken powerlaw
below 10~Hz and a QPO is apparent above 10~Hz.  Fitting the power
spectrum with the sum of a broken powerlaw plus a Lorentzian, we found
excess noise above 20~Hz and $\chi^2/\rm DoF = 143.5/75$.  We added an
exponentially cutoff powerlaw to fit this excess noise and the fit
improved to $\chi^2/\rm DoF = 111.8/72$.  The residuals of this fit
show no systematic deviations and the fit is unlikely to be improved by
addition of more continuum components.  For the broken powerlaw, we
find a break frequency of $5.7 \pm 0.2 \rm \, Hz$, an index below the
break of $-0.25 \pm 0.02$ and and index above the break of $-1.7 \pm
0.2$.  The Lorentzian centroid is $17.7 \pm 0.2 \rm \, Hz$.  The break
and QPO centroid frequencies are consistent with the correlation
reported by \citet{wijnands99}.

The power spectrum for April 12 and 15 shows weak timing noise.  The
RMS fraction is 1.4\% in the 0.1--100~Hz range.  This indicates that
the source was in the `banana' state.  We fitted the power spectrum
with the sum of a powerlaw and an exponentially cutoff powerlaw.  This
gave a reasonable fit with $\chi^2/\rm DoF = 11.6/10$.  The powerlaw
index was $-0.89 \pm 0.14$.

Based on the timing and color information, we suggest that SAX
J1750.8-2900 is an atoll source.  Furthermore, we identify the source
as being in the ``island'' state on April 6 and 9, when the hard color
had high values and strong timing noise was present, and in the
``banana'' state on April 12 and 15, when the hard color had low values
and the timing noise was weak.

\begin{deluxetable}{lcccccc}
\tablecolumns{10}
\tablecaption{High frequency peaks in the persistent emission
  \label{qpotable}}
\tablewidth{0pt}
\tablehead{
  \colhead{Time}  & \colhead{Duration} & \colhead{Hard color} &
  \colhead{Centroid} & \colhead{Width} & \colhead{Amplitude} \\ 
  \colhead{(UTC)} & \colhead{(s)}      &                      &
  \colhead{(Hz)}     & \colhead{(Hz)}  & \colhead{(\%)} \\
    }
\startdata
Apr 06 at 13:03:27 & 1344 & 0.46 & $543\pm 23$ & $210\pm 70$ & $8.8\pm 2.1$ \\
Apr 09 at 10:59:27 & 3024 & 0.45 & $651\pm 15$ & $228\pm 45$ & $9.5\pm 1.3$ \\
Apr 12 at 13:41:03 & 2336 & 0.37 & $1017\pm 4$ & $50\pm 12$  & $2.9\pm 0.5$ \\
Apr 15 at 14:57:35 & 2960 & 0.35 & $936\pm 1$  & $8\pm 2$    & $0.67 \pm 0.13$ \\
Apr 15 all data    & 8608 & 0.35 & $1253\pm 9$ & $60\pm 24$  & $0.9\pm 0.2$ \\
\enddata

\tablecomments{The table includes: Time -- the UTC time at the
beginning of the observation, all observations were in 2001; Duration
-- of the observation; Hard color --  as defined in Fig.~\ref{ccd};
Centroid and Width - of the fitted Lorentzian; Amplitude - RMS fraction
of QPO signal.}   \end{deluxetable}

\begin{figure}[tb] \epsscale{1.0} \plottwo{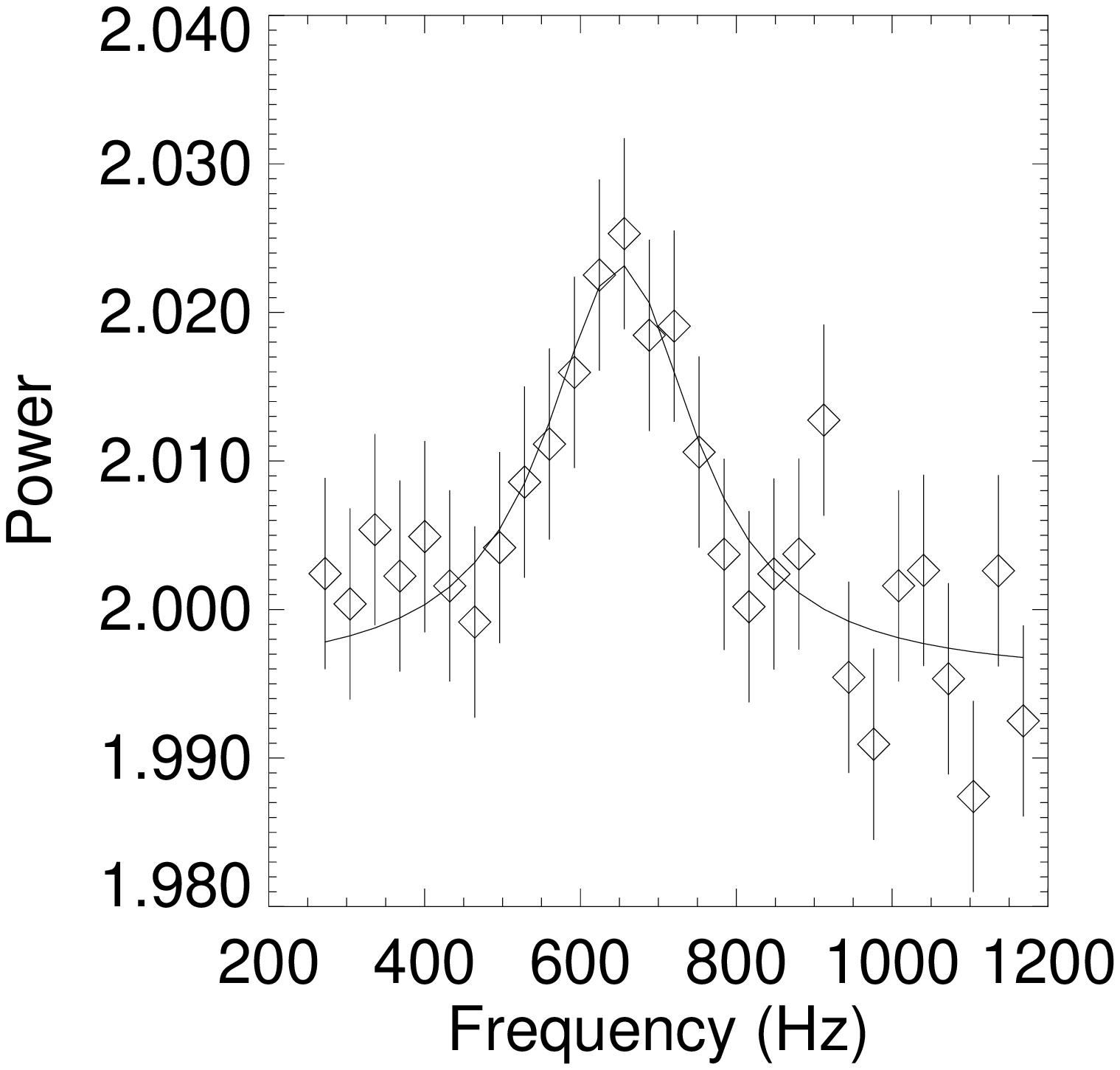}{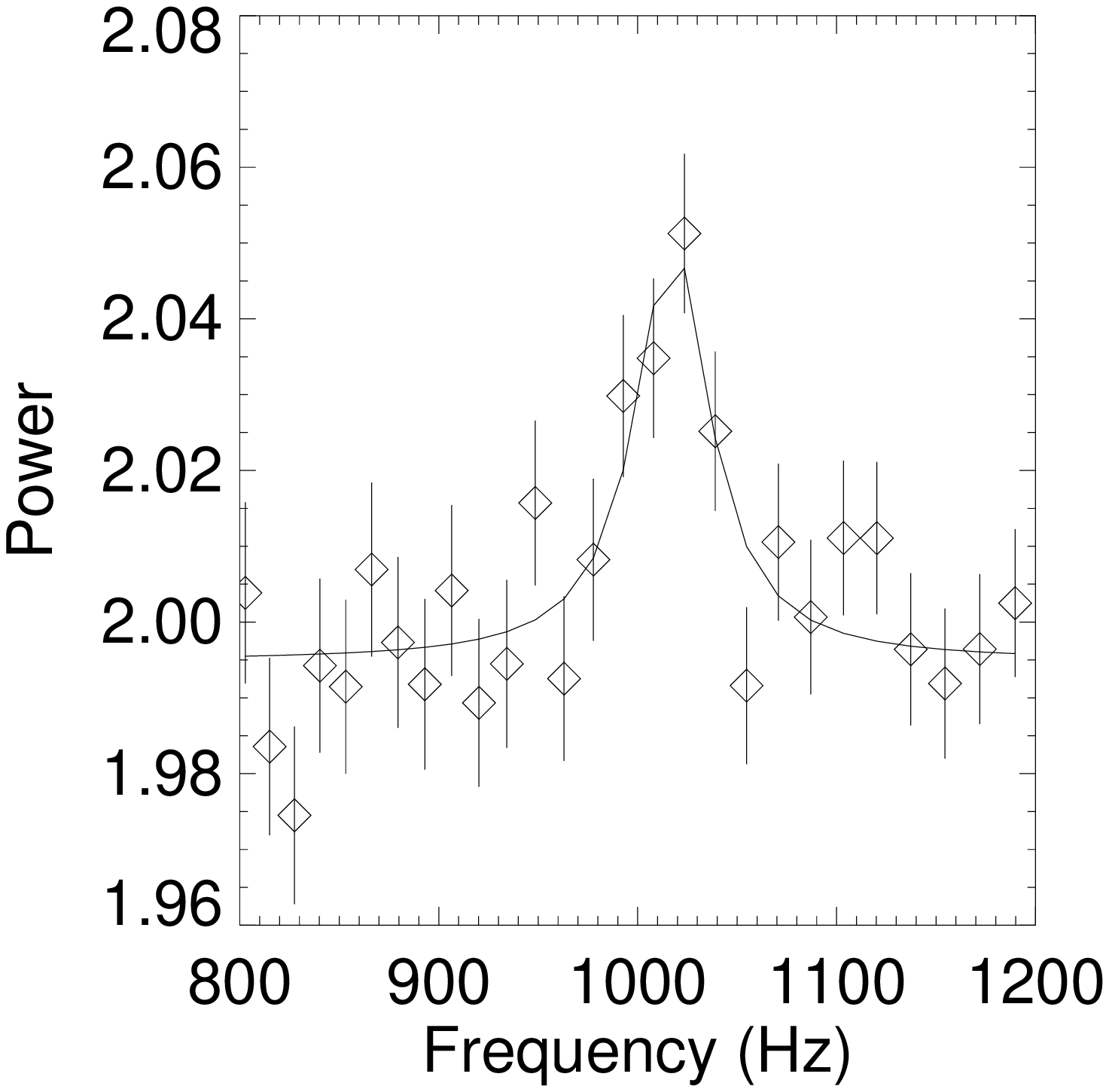}
\caption{Power spectra showing kHz QPOs.  The panel on the left uses
data from April 9, on the right from April 12.  The power spectra were
calculated using events from the full PCA band (2--60~keV).  The power
is Leahy normalized.} \label{khzqpo} \end{figure}

\subsection{High frequency timing} 

We searched for high frequency QPOs in each uninterrupted RXTE
observation window and in combinations of the various data segments. 
We calculated averages of 2~s power spectra for all PCA events
(2--60~keV) and for events in the 4.7-20.8~keV energy band.  We
included events from all PCUs on during each observation.  We note that
the true energy band for PCU0 likely differs somewhat from the nominal
range due to the propane layer leak.  We searched for peaks in the
spectrum and fit any peak found with a Lorentzian plus a constant equal
to the calculated the Poisson noise level.

There were several QPO signals above 250~Hz, see Table~\ref{qpotable}. 
Those found on April 9 and 12 are strong detections and are shown in 
Fig.~\ref{khzqpo}.  Allowing for a number of trials equal to the search
interval of 1200~Hz divided by the QPO width \citep{vStra00} and
allowing several trial widths, we estimate a chance probability of
occurrence of $2 \times 10^{-11}$ for the QPO on April 9 and  $3 \times
10^{-6}$ for the QPO on April 12.  These QPOs  clearly establish SAX
J1750.8-2900 as a new member of the class of neutron-star low-mass
X-ray binaries (NS-LMXBs) exhibiting kHz QPOs.   The other signals have
lower significance: $2 \times 10^{-3}$ for the QPO on April 6 and $2
\times 10^{-4}$ for the QPO on April 15.  The signal on April 15 is
quite narrow and appears significant only in the 4.7-20.8~keV band
power spectrum.  

In addition to the signals detected in individual uninterrupted
observation segments, we also found a signal at 1253~Hz in the sum of
all of the April 15 data.  This is the last entry in
Table~\ref{qpotable}.  The detection has relatively low significance
and must be considered tentative.  However, if the detection is
correct, then the frequency difference between the two QPOs detected on
April 15 is 317~Hz.  The formal error on the difference of the two
frequency centroids is 9~Hz.  The 1253~Hz peak may be broadened due to
shifts in the centroid frequency over the integration, so the true
uncertainty on the difference is somewhat larger.  The frequency
difference is consistent with half the frequency of the burst
oscillations.

\section{Discussion}

The results presented here establish that SAX J1750.8-2900  exhibits
millisecond oscillations in its X-ray emission.  The properties of the
kHz QPOs in the persistent emission from SAX J1750.8-2900 are similar
to those of the other neutron-star LMXB kHz QPO sources in terms of the
observed frequency range and the oscillation amplitudes.  Additional
measurements of SAX J1750.8-2900 would be of interest to confirm our
tentative detection of the second kHz QPO branch and obtain a
simultaneous measurement of the kHz QPO frequency difference.

The properties of the burst oscillations from SAX J1750.8-2900 are also
similar to those observed from other sources.  The frequency shift seen
in the single burst from SAX J1750.8-2900 exhibiting oscillations is
compatible with those seen from several other sources and smaller than
the large shift (by 1.32\%) seen from 4U~1916-053 \citep{galloway01}.

X-ray burst oscillation sources appear to form two distinct classes:
``fast oscillators'' showing burst oscillations near 600~Hz with the
precise frequency close to twice the frequency difference of the kHz
QPOs seen in the persistent emission, and ``slow oscillators'' showing
burst oscillations near 300~Hz with the precise frequency near the
difference of the kHz QPO frequencies \citep{white97}.  Our discovery
of 600.75~Hz oscillations from SAX J1750.8-2900 establish that it is a
member of the class of fast X-ray burst oscillators. Our tentative
detection of two contemporaneous kHz QPOs with a frequency difference
near 300~Hz is consistent with the properties of the other fast
oscillators, but should be tested with additional observations.

\citet{muno01} showed that the fast oscillators produce burst
oscillations predominately, but not exclusively, in photospheric radius
expansion bursts, while the slow oscillators have oscillations in
bursts both with and without photospheric radius expansion.  The burst
from SAX J1750.8-2900 in which we find oscillations shows weak, at
best, evidence for photospheric radius expansion; in particular, it
does not meet the requirement of 20~km of radius expansion of
\citet{muno01}.  However, since we have a sample of only one burst with
oscillations, our results are not inconsistent with those of
\citet{muno01}.

The fact that the behavior of the fast versus slow oscillators is
different in regards to the occurrence of oscillations versus burst
type (radius expansion or not) implies that the difference between the
two classes is not an observational selection effect, i.e.\ due to  the
inclination of the neutron star spin axis relative to our line of
sight, but rather a physical difference in the properties of the
neutron stars \citep{muno01}.  It would be of great interest to
identify a model of the X-ray burst oscillations which explained both
the large frequency shifts seen in some bursts \citep{galloway01} and
the dichotomy between the fast versus slow oscillators.

\acknowledgments  

We greatly appreciate the assistance by the duty scientists of the
BeppoSAX Science Operations Center in the near to real-time WFC data
analysis.  We gratefully acknowledge the efforts of the RXTE team,
particularly Jean Swank and Evan Smith, in performing these target of
opportunity observations.  PK thanks Mal Ruderman for useful
discussions and acknowledges support from NASA grants NAG5-7405,
NAG5-9097, and NAG5-9104.  JZ acknowledges financial support from the
Netherlands Organization for Scientific Research (NWO).



\end{document}